\documentclass{llncs}

  \usepackage{amsmath, amssymb, latexsym}
  \usepackage{graphicx}
\usepackage{epic,eepic}

\usepackage{psfrag}

\usepackage{url}
\usepackage{amssymb,epsfig,amstext}
\usepackage{txfonts}
\usepackage{algorithmic}
\usepackage{algorithm}
\usepackage{graphicx}
\usepackage{multirow}
\usepackage{float}
\usepackage{picinpar}
\usepackage{listings}

\newcommand{\AP}{\emph{AP}}

\newcommand{\G}{\textbf{G}}

\newcommand{\F}{\textbf{F}}
\newcommand{\Next}{\textbf{X}}

\newcommand{\AAA}{\mathcal{A}}

\newcommand{\U}{\textbf{U}}

\newcommand{\PP}{\mathcal{P}}
\newcommand{\MM}{\mathcal{M}}

\newcommand{\CC}{\mathcal{D}}
\newcommand{\Conf}{\mathcal{C}}

\newcommand{\mg}{\mathcal{G}}

\newcommand{\Lang}{\mathbb{L}}

\newcommand{\tl}{\mathbb{f}}

\newcommand{\BA}{\mathcal{B}}

\def\longleadsto#1{\mathop{{\hbox{\setbox0=\hbox{$\scriptstyle{#1\quad}$}{$%
\mathrel{\mathop{\setbox1=\hbox to
\wd0{\leadstofill}\ht1=3pt\dp1=-2pt\box1}\limits^{#1}}%
$}}}}}

\def\by#1{\mathop{{\hbox{\setbox0=\hbox{$\scriptstyle{#1\quad}$}{$%
\mathrel{\mathop{\setbox1=\hbox to
\wd0{\rightarrowfill}\ht1=3pt\dp1=-2pt\box1}\limits^{#1}}%
$}}}}}

\def\byi#1{\mathop{{\hbox{\setbox0=\hbox{$\scriptstyle{#1\quad}$}{$%
\mathrel{\mathop{\setbox1=\hbox to
\wd0{\rightarrowfill$_i$}\ht1=3pt\dp1=-2pt\box1}\limits^{#1}}%
$}}}}}

\def\bystar#1{\mathop{{\hbox{\setbox0=\hbox{$\scriptstyle{#1\quad}$}{$%
\mathrel{\mathop{\setbox1=\hbox to
\wd0{\rightarrowfill$^*$}\ht1=3pt\dp1=-2pt\box1}\limits^{#1}}%
$}}}}}

\def\2by#1{\mathop{{\hbox{\setbox0=\hbox{$\scriptstyle{#1\quad}$}{$%
\mathrel{\mathop{\setbox1=\hbox to
\wd0{\Rightarrowfill}\ht1=3pt\dp1=-2pt\box1}\limits^{#1}}%
$}}}}}

        {\hspace*{\fill}$\Box$}

\def\myautorightarrow#1{\mathop{{\hbox{\setbox0=\hbox{$\scriptstyle{#1\quad}$}{$%
\mathrel{\mathop{\setbox1=\hbox to
\wd0{\mapstofill}\ht1=3pt\dp1=-2pt\box1}\limits^{#1}}%
$}}}}}

\title{LTL Model-Checking for Dynamic Pushdown Networks Communicating via Locks}

\author{ Fu Song\inst{1}  \and  Tayssir Touili\inst{2}}

\institute{Shanghai Key Laboratory of Trustworthy Computing, East China Normal University, Shanghai, China
\and
LIPN, CNRS and University Paris 13, France}

\begin{document}

\maketitle

\begin{abstract}
A Dynamic Pushdown Network (DPN) is a set of pushdown systems (PDSs) where each process can dynamically create new instances of PDSs.
DPNs are a natural model of multi-threaded programs with (possibly recursive) procedure calls and thread creation. Extending DPNs with locks allows processes to synchronize with each other. Thus, DPNs with locks are a well adapted formalism to model multi-threaded programs that
synchronize via locks. Therefore, it is important to have model-checking algorithms for DPNs with locks.
We consider in this work model-checking for DPNs with locks against single-indexed LTL properties of the form $\bigwedge f_i$
s.t. $f_i$ is a LTL formula interpreted over the PDS $i$. We consider the model-checking problems w.r.t. simple valuations (i.e, whether a configuration satisfies an atomic proposition depends only on its control location and held locks) and w.r.t. regular valuations (i.e., the set of the configurations satisfying an atomic proposition is a regular set of configurations). We show that these model-checking problems are decidable. 
\end{abstract}

\section{Introduction}
It is notoriously difficult to write multithreaded programs  whose bugs are commonly concurrency-related and
are hard to reproduce and fix.  Dynamic Pushdown Networks (DPN) \cite{BOT05} are a natural model of multi-threaded programs with (possibly recursive) procedure calls and thread creation. A DPN consists of a finite set of pushdown systems (PDSs), each of them models a sequential program (process) that can dynamically create new instances of PDSs. The model-checking programs (LTL, CTL or reachability properties) of DPNs
are well studied in the literature \cite{BOT05,Lug11,LOW09,GLOSW11,LO07,Wen10,ST13dpn}. DPNs with locks is an extension of DPNs in which processes can synchronize with each other via locks. DPNs with locks allow to model multi-threaded programs communicating via locks.
However, only reachability properties are studied for DPNs with locks \cite{LOW09,GLOSW11}. In this work, we consider model checking Linear Temporal Logic (LTL) which can describe more interesting properties of program behaviors. 

In general, model checking DPNs for double-indexed properties (i.e., properties where atomic propositions are interpreted over the control states of two or more threads) or model checking DPNs with locks for reachabillity properties is undecidable \cite{KG06}. This undecidability holds for pushdown networks even without thread creation. To obtain decidable results, in this paper, we consider single-indexed LTL properties for DPNs using locks in a nested style (L-DPN), where a single-index LTL formula is a formula of the form $\bigwedge f_i$ such that $f_i$ is a LTL formula interpreted over the PDS $i$, and using locks in a nested style allows each process to release only the latest acquired lock
that has not yet been released.   A L-DPN satisfies $\bigwedge f_i$ iff every PDS $i$ running in parallel in the network satisfies the subformula $f_i$ and the lock criteria \cite{LOW09}.

We consider single-indexed LTL model-checking for L-DPNs with simple valuations (where whether a configuration of a PDS $i$ satisfies
an atomic proposition depends only on the control location of the configuration and its held locks) and with regular valuations (where the set of configurations of a PDS satisfying an atomic proposition is a regular set of configurations).
We show that these model-checking problems are decidable.
It is non-trivial to do LTL model checking for L-DPNs, since the number of instances of PDSs can be unbounded. Checking independently whether all the different PDSs satisfy the corresponding subformula $f_i$ is not correct. Indeed, we do not need to check whether an instance of a PDS $j$ satisfies $f_j$ if this instance is not created during a run, and we have to guarantee that all the created instances use locks in a nested style. In our previous work \cite{ST13dpn}, we have shown how to solve single-indexed LTL model-checking for L-DPNs without locks, i.e., DPNs.
However, the approach of \cite{ST13dpn} cannot be directly applied to perform single-indexed LTL model-checking for L-DPNs due to locks.
Indeed, we have to consider communication between each instance running in parallel in the network. To overcome this problem,
inspired by the work of \cite{LMW09} which reduces reachability checking of L-DPNs to checking reachability of DPNS,
we reduce single-indexed LTL model-checking for L-DPNs to single-indexed LTL model-checking for DPNs.
For this, we will compute a DPN, a kind of ``product" of the L-DPN with acquisition structures, where each control location of the DPN
stores an acquisition structure. The acquisition structures stored in the control locations allow us to infer whether the global run uses locks correctly. For this, we characterize the set of consistent acquisition structures that will not violate the lock usages.
The transformation of the acquisition structure during the global run of the DPN
checks whether the lock usages is violated or not (i.e., the acquisition structure is consistent) and the runs of the DPN mimics the global run of the L-DPN. We disallow the global runs of the DPN in which an inconsistent acquisition structure will meet.
By doing this, the global runs of the DPN exactly correspond to the global runs of the L-DPN that uses locks correctly.
Thus, we reduce single-indexed LTL model-checking for L-DPNs
to single-indexed LTL model-checking for DPNs. This later problem can be solved by our previous work \cite{ST13dpn}.

\medskip
\noindent
{\bf Outline.} Section \ref{prel} gives the basic definitions, recalls the results of single-indexed LTL model-checking for DPNs and
presents a motivating example. Section \ref{lock-elm} shows how to solve single-indexed LTL model-checking for L-DPNs.
Section \ref{rel} discusses related works.

\section{Preliminaries}
\label{prel}

\subsection{LTL and B\"{u}chi Automata}
From now on, we fix a set of atomic propositions $\AP$.
\begin{definition}The set of LTL formulas is given by (where $ap\in \AP$):

\hspace*{30mm}$\psi~::= ap~\mid~\neg \psi~\mid~\psi\wedge\psi~\mid~\Next\psi~\mid~\psi\U\psi.$
\end{definition}

Given an $\omega$-word $\eta=\alpha_0\alpha_1...$ over $2^\AP$, let $\eta(k)$ denote $\alpha_k$, and $\eta_k$ denote the \emph{suffix} of $\eta$ starting from
$\alpha_k$. $\eta\models\psi$ ($\eta$ satisfies $\psi$) is inductively defined as follows:
$\eta\models ap$ iff $ap\in\eta(0)$; $\eta\models \neg\psi$ iff $\eta\not\models \psi$; $\eta\models \psi_1\wedge\psi_2$ iff $\eta\models \psi_1$ and $\eta\models \psi_2$; $\eta\models \Next\psi$ iff $\eta_1\models \psi$; $\eta\models \psi_1\U\psi_2$ iff there exists $k\geq 0$ such that $\eta_k\models \psi_2$ and
for every $j$, $1\leq j<k$, $\eta_j\models \psi_1$.

\begin{definition}
A \emph{B\"{u}chi automaton} ($BA$) $\BA$ is a tuple $(G,\Sigma,\theta, g^0, F)$ where
$G$ is a finite set of states, $\Sigma$ is the input alphabet, $\theta\subseteq G\times\Sigma \times G$
is a finite set of transitions, $g^0\in G$ is the initial state and $F\subseteq G$ is a finite set of accepting states.
\end{definition}
A run of $\BA$ over an $\omega$-word $\alpha_0\alpha_1...$ is a sequence of states $q_0q_1...$ s.t. $q_0=g^0$ and $(q_i,\alpha_i,q_{i+1})\in\theta$ for every
$i\geq 0$. A run is accepting iff it infinitely often visits some states in $F$.

It is well-known that given a LTL formula $f$, one can construct a BA $B_f$ s.t. $\Sigma=2^{AP}$ recognizing all the $\omega$-words that satisfy $f$ \cite{VW86}.

\subsection{Dynamic Pushdown Networks with Locks}

\begin{definition}
A \emph{Dynamic Pushdown Network} with Locks (L-DPN) $\MM$ is a tuple $(Act,\Lang, \PP_1,...,\PP_n)$ s.t. $\Lang$ is a finite set of locks, $Act$ is a finite set of actions $\{acq(l),rel(l),\tau\mid l\in \Lang\}$ where the action $acq(l)$ (resp. $rel(l)$) for every $l\in \Lang$
denotes the \emph{acquisition} (resp. \emph{release}) of the lock $l$ and the action $\tau$ denotes all the lock-unrelated \emph{internal actions};  for every $i$,
$1\leq i\leq n$, $\PP_i=(P_i,\Gamma_i,\Delta_i)$ is a
\emph{Dynamic Pushdown System} (DPDS), where $P_i$ is a finite set of control
states s.t. $P_k\cap P_i=\emptyset$ for $k\neq i$,  $\Gamma_i$ is the stack alphabet, $\Delta_i$ is a finite set of transition rules in the following forms: (I)
$p_0\gamma\stackrel{a}{\hookrightarrow}_i p_1\omega_1$ or (II)
$p_0\gamma\stackrel{a}{\hookrightarrow}_i p_1\omega_1\rhd p_2\omega_2$ s.t.
$a\in Act, p_0,p_1\in P_i,\gamma\in\Gamma_i,\omega_1\in\Gamma_i^*,p_2\omega_2\in
P_j\times\Gamma_j^*$ for some $j,~1\leq j\leq n$.

\medskip
A L-DPN $\MM$ is a Dynamic Pushdown Network (DPN) if $Act=\{\tau\}$ and $\Lang=\emptyset$. We will write a DPN
as $(\PP_1,...,\PP_n)$ and sometimes we omit the labeling $\tau$ from all the transition rules in DPNs.
 \end{definition}

For every $i: 1\leq i\leq n$, a \emph{local configuration} of a DPDS $\PP_i$ is a tuple $(p\omega, L)$ such that
$L\subseteq \Lang$ is a set of held locks,
$p\in P_i$ is the control location and $\omega\in\Gamma_i^*$ is the stack content. Note that, if $Act=\{\tau\}$ and $\Lang=\emptyset$,  a local configuration is denoted by $p\omega$. A DPDS is a pushdown system if all the transition rules are in the form of $q\gamma\stackrel{a}{\hookrightarrow}_i p_1\omega_1$.

A \emph{global configuration} of $\MM$ is a multiset over $\bigcup_{i=1}^nP_i\times\Gamma_i^*\times 2^\Lang$, in which
each element $(p\omega, L)$ denotes the local configuration of an instance running in parallel in the network.
Given a global configuration $\mg$,
the set of held locks at the global configuration is $(\bigcup_{(p\omega,L)\in\mg} L)$, denoted by $hold(\mg)$.
The set of free locks at the global configuration $\mg$ is $\Lang\setminus hold(\mg)$, denoted by $free(\mg)$.
Let $\Conf_\MM$ denote the set of global configurations of $\MM$.

Given a tree $T$ over $(\bigcup_{i=1}^n P_i\times\Gamma_i^*)\times2^\Lang$, a node in the tree $T$ that does not have any child is a leaf.
Let $leaves(T)$ be the multiset that contains exactly the leaves of the tree $T$. W.l.o.g., we assume that
the initial global configuration $\mg$ of $\MM$ contains only one element.
A \emph{global run} of $\MM$ starting from an initial global configuration $\mg$ containing the element $(p_0\omega_0,L_0)$ is a binary tree $T$ rooted by $(p_0\omega_0,L_0)$ and the leaves $leaves(T)$ of $T$ is the current global configuration. The progress of the global run $T$ is defined as follows: for every local configuration $(p\gamma u, L)\in leaves(T)$ of an instance of a DPDS $\PP_i$ running in parallel in the network for some $\gamma\in \Gamma_i$:

 \begin{itemize}
   \item[$\alpha_1:$] If there exists a transition rule $p\gamma\stackrel{\tau}{\hookrightarrow}_i p'\omega\in\Delta_i$, then $(p'\omega u, L)$ can be the right child of the node $(p\gamma u, L)$. This means that this instance can move from $(p\gamma u, L)$ to $(p'\omega u, L)$, replacing  the control location $p$ by $p'$ and the stack content $\gamma u$ by $\omega u$ without changing the set of held locks $L$. During this step, the other local configurations $leaves(T)\setminus(p\gamma u,L)$ of other instances running in parallel in the network stay at the same local configurations. After applying this transition rule, the next global configuration is $\{(p'\omega u,L)\}\cup ( leaves(T)\setminus(p\gamma u,L))$.

   \item[$\alpha_2:$] If there exists a transition rule $p\gamma\stackrel{\tau}{\hookrightarrow}_i p'\omega\rhd p_2\omega_2\in\Delta_i$ s.t.
   $p_2\in P_j$, then $(p'\omega u, L)$ and  $(p_2\omega_2, \emptyset)$ can be
   the right child and left child of the node $(p\gamma u, L)$. This means that this instance can move from $(p\gamma u, L)$ to $(p'\omega u, L)$. Moreover, a new instance of the DPDS $\PP_j$ is created and it starts from the local configuration $(p_2\omega_2,\emptyset)$. Here, we suppose w.l.o.g., that the set of locks held by this new instance is empty. This kind of newly created instance during the global run is \emph{called dynamically created new instance}, and a local configuration that a dynamically created new instance starts from is called by \emph{Dynamically Created Local Initial Configuration} (DCLIC for short).
   The other local configurations $leaves(T)\setminus(p\gamma u,L)$ of other instances running in parallel in the network stay at the same local configurations. After applying this transition rule, the next global configuration is $\{(p'\omega u,L),(p_2\omega_2,\emptyset)\}\cup (leaves(T)\setminus(p\gamma u,L))$.

   \item[$\alpha_3:$] If there exists a transition rule $p\gamma\stackrel{acq(l)}{\hookrightarrow}_i p'\omega\in\Delta_i$ such that
   $l\in free(leaves(T))$ (i.e., $l$ is a free lock at the current global configuration $leaves(T)$), then $(p'\omega u, L\cup\{l\})$ can be
   the right child of the node $(p\gamma u, L)$. This means that this instance can move from $(p\gamma u, L)$ to $(p'\omega u, L\cup\{l\})$ and hold the lock $l$. During this step, the other local configurations $leaves(T)\setminus(p\gamma u,L)$ of other instances running in parallel in the network stay at the same local configurations. After applying this transition rule, the next global configuration is $\{(p'\omega u,L\cup\{l\})\}\cup( leaves(T)\setminus(p\gamma u,L))$.

   \item[$\alpha_4:$] If there exists a transition rule $p\gamma\stackrel{acq(l)}{\hookrightarrow}_i p'\omega\rhd p_2\omega_2\in\Delta_i$ such that
   $l\in free(leaves(T))$, then $(p'\omega u, L\cup\{l\})$ and $(p_2\omega_2, \emptyset)$ can be
   the right child and left child of the node $(p\gamma u, L)$. This means that this instance can move from $(p\gamma u, L)$ to $(p'\omega u, L\cup\{l\})$, hold the lock $l$ and create a new instance starting from $(p_2\omega_2,\emptyset)$.  During this step, the other local configurations $leaves(T)\setminus(p\gamma u,L)$ of other instances running in parallel in the network stay at the same local configurations. After applying this transition rule, the next global configuration is $\{(p'\omega u,L\cup\{l\}), (p_2\omega_2,\emptyset)\}\cup( leaves(T)\setminus(p\gamma u,L))$.

   \item[$\alpha_5:$] If there exists a transition rule $p\gamma\stackrel{rel(l)}{\hookrightarrow}_i p'\omega\in\Delta_i$ such that
   $l\in L$ (i.e., the instance owns the lock $l$), then $(p'\omega u, L\setminus\{l\})$ can be
   the right child of the node $(p\gamma u, L)$. This means that this instance can move from $(p\gamma u, L)$ to $(p'\omega u, L\setminus\{l\})$ and free the lock $l$. The other local configurations $leaves(T)\setminus(p\gamma u,L)$ of other instances running in parallel in the network stay at the same local configurations. After applying this transition rule, the next global configuration is $\{(p'\omega u,L\setminus\{l\})\}\cup( leaves(T)\setminus(p\gamma u,L))$.

   \item[$\alpha_6:$] If there exists a transition rule $p\gamma\stackrel{rel(l)}{\hookrightarrow}_i p'\omega\rhd p_2\omega_2\in\Delta_i$ such that
   $l\in L$ (i.e., the instance owns the lock $l$), then $(p'\omega u, L\setminus\{l\})$ can be
   the right child of the node $(p\gamma u, L)$. This means that this instance can move from $(p\gamma u, L)$ to $(p'\omega u, L\setminus\{l\})$, free the lock $l$ and create a new instance starting from $(p_2\omega_2,\emptyset)$. The other local configurations $leaves(T)\setminus(p\gamma u,L)$ of other instances running in parallel in the network stay at the same local configurations. After applying this transition rule, the next global configuration is $\{(p'\omega u,L\setminus\{l\}),(p_2\omega_2,\emptyset)\}\cup( leaves(T)\setminus(p\gamma u,L))$.
    \end{itemize}

Intuitively, each left child in a global run $T$ is a local initial configuration of the newly created instance and the root
is the local initial configuration of the initial instance (i.e., the instance is not created on the runtime).
Each rightmost path is a trance of an instance running in parallel in the global run (network). The rightmost path
starting from the root or a left child is a \emph{local run} of the initial instance or a newly created instance
running in parallel in the network, respectively. Note the defining the global runs as trees
allows us to know which instance creates a new instance, where it creates and the local initial configuration
of the newly created instance. It is important to reasoning about lock usages.
Let $\CC_I=\{(p_2\omega_2,\emptyset)\mid p\gamma\stackrel{a}{\hookrightarrow}_i p'\omega\rhd p_2\omega_2\in\Delta_i \mbox{ for some }
i, 1\leq i\leq n\}$ be the set of all the possbile DCLICs. When the L-DPN is DPN,
$\CC_I=\{p_2\omega_2\mid p\gamma\stackrel{a}{\hookrightarrow}_i p'\omega\rhd p_2\omega_2\in\Delta_i \mbox{ for some }
i, 1\leq i\leq n\}$.

\medskip
\noindent{\bf Nested Lock Access}.
A global run of $\MM$ uses locks in a nested style iff each local run running in parallel in the global run
uses locks in a nested style, i.e., the local run releases only the latest acquired lock that has not yet
been released. In this work, we consider L-DPNs that use locks in a nested style.
This is because even reachability, and hence LTL, is known to be
undecidable for pushdown networks that use locks in an arbitary style \cite{KG06}.

\begin{figure}[t]
\centering
  \includegraphics[width=0.9\textwidth]{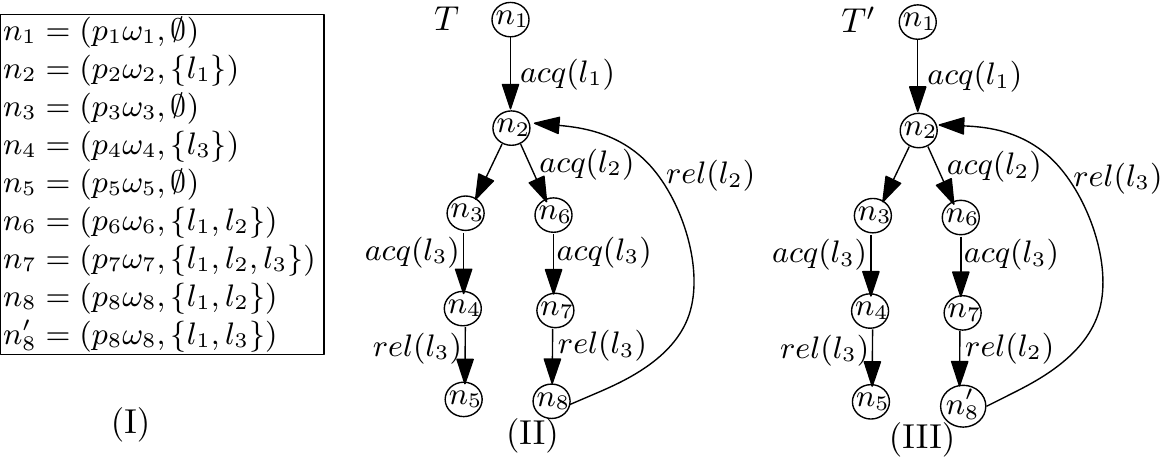}
  \caption{(a) and (b) are two global runs that using locks (not) in a nested style.}\label{nested-locks}
\end{figure}

\begin{example}
Figure \ref{nested-locks}(II) and Figure \ref{nested-locks}(III) show two global runs called by $T$ and $T'$ that uses locks $l_1$, $l_2$ and $l_2$. Each edge is labeled by the corresponding action. The nodes $n_1,...,n_8$ and $n_8'$ denote the local configurations showing in Figure \ref{nested-locks}(I).
$n_1(n_2n_6n_7n_8)^*$ (resp. $n_1(n_2n_6n_7n_8')^*$) is a local run in $T$ (resp. $T'$) during which a new instance is created
when moving from $n_2$ to $n_6$. This new instance has the local run $n_3n_4n_5$.
In $T$, we can see that locks are accessed in a nested style. While, in $T'$, the locks $l_1$ and $l_2$ are not accessed in a nested style,
since the lock $l_2$ is released before the release of the latest acquired lock $l_3$.
\end{example}

W.l.o.g., in this work, we consider only non-reentrant locks, i.e., a process cannot acquire the same lock multiple times before releasing
it. While reentrant locks allow a process to acquire the same lock multiple times before releasing it.
Indeed, reentrant locks can be simulated with non-reentrant locks \cite{LOW09}.

\subsection{Representing Infinite Set of Local Configurations}
To finitely represent (infinite) sets of local configurations of DPDSs and DCLICs generated by DPDSs,
we introduce L-Multi-automata and Multi-automata.

\begin{definition}
Given a L-DPN $\MM=(Act,\Lang,\PP_1,...,\PP_n)$,
a L-Multi-automaton {\em (L-MA)} is a tuple $\AAA_i=(Q_i,\Gamma_i,\delta_i,I_i,Acc_i)$, where $Q_i$ is a finite set of states, $I_i\subseteq P_i\times 2^\Lang$ is a finite set of initial states corresponding to the control locations and held locks
of the DPDS $\PP_i$, $Acc_i\subseteq Q_i$ is a finite set of final states, $\delta_i\subseteq (Q_i\times\Gamma_i)\times 2^{\CC_I}\times Q_i$ is a finite set of transition rules.
\medskip
A L-MA is a Multi-automaton (MA) if the L-DPN is a DPN, i.e., $I_i\subseteq P_i$ (note that $\Lang=\emptyset$).

\end{definition}
We write  $q\by{\gamma/D}_i~q'$ instead of $(q,\gamma,D,q')\in \delta_i$, where $D$ is a set of DCLICs. We define the relation $\longrightarrow_i^*\subseteq(Q_i\times\Gamma_i^*)\times 2^{\CC_I}\times Q_i$ as the smallest relation s.t.:
(1) $q{\bystar{\epsilon/\emptyset}\hspace*{-1.5mm}_i}~q$ for every $q\in
Q_i$, (2) if $q\by{\gamma/D_1}_i~q_1$ and
$q_1{\bystar{\omega/D_2}\hspace*{-1.5mm}_i}~q_2$, then
$q{\bystar{\gamma\omega/D_1\cup D_2}\hspace*{-1.5mm}_i}~q_2$. Let $L(\AAA_i)$ be the set of
tuples $(p\omega,L,D)\in P_i\times\Gamma_i^*\times 2^{\CC_I}$ s.t. $(p,L){\bystar{\omega/D}\hspace*{-1.5mm}_{i}}~q$ for some $q\in Acc_i$.
A set $W\subseteq P_i\times\Gamma_i^*\times 2^\Lang\times 2^{\CC_I}$ is \emph{regular} iff there exists a L-MA $\AAA_i$ s.t. $L(\AAA_i)=W$.
A set of local configurations $C \subseteq P_i\times\Gamma_i^*\times 2^\Lang$ is \emph{regular} iff $C\times\{\emptyset\}$ is a regular set.

\subsection{Single-indexed LTL Model-Checking for L-DPNs}
Model-checking L-DPNs for double-indexed LTL properties where the validity of atomic propositions
depends on two or more DPDSs is undecidable. Indeed, model-checking double-indexed LTL properties for
pushdown networks even without interaction with each other is undecidable \cite{KG06}.
Thus, in this work, we consider model-checking L-DPNs for single-indexed LTL properties of the form $\bigwedge_{i=1}^n f_i$
s.t. for every $i: 1\leq i\leq n$, $f_i$ is a LTL formula interpreted over the DPDS $\PP_i$.  From now on, we fix a L-DPN $\MM=(Act,\Lang,\PP_1,...,\PP_n$ s.t. for every $i$,
$1\leq i\leq n$, $\PP_i=(P_i,\Gamma_i,\Delta_i)$ and a single-index LTL formula $\bigwedge_{i=1}^n f_i$ such that
for every $i:1\leq i\leq n$, $f_i$ is  interpreted over the DPDS $\PP_i$.

Given a valuation $\lambda:  AP\longrightarrow 2^{\bigcup_{i=1}^n (P_i\times\Gamma_i^*\times 2^\Lang)}$ that assigns to each atomic proposition
a set of local configurations, and a global run $T$ of the L-DPN $\MM$,
a local run $\sigma=(p_0\omega_0,L_0)(p_1\omega_1,L_1)...$ of an instance of $\PP_i$ running in parallel in the global run $T$ satisfies $f_i$ iff the $\omega$-word $\sigma_0\sigma_1...$ satisfies $f_i$, where for every $j\geq 0$, $\sigma_j=\{ap\in AP \mid (p_j\omega_j,L_j)\in\lambda(at)$.
A global run $T$ of the L-DPN $\MM$ satisfies $f$ iff each local run of each instance of $\PP_i$ for $1\leq i\leq n$ running in parallel in the network satisfies $f_i$. A global initial configuration $\mg$ satisfies $f$ iff $\MM$ has a global run $T$ starting from $\mg$ such that $T$ satisfies $f$.

However, checking whether a local run of an instance of a pushdown system $\PP_i$ without locks satisfies the formula $f_i$ with respect to an arbitary valuation $\lambda$ is undecidable \cite{esparza-ctl*1}.  To have decidable results, in this work, we first consider
L-DPNs model-checking for single-indexed LTL with simple valuations, where the validity of each atomic proposition
only depends on the control location and locks. Formally, a simple valuation is a function $\lambda:  AP\longrightarrow 2^{\bigcup_{i=1}^n (P_i\times 2^\Lang)}$ that assigns to each atomic proposition a set of control locations and locks. It can be extended
to $\lambda:  AP\longrightarrow 2^{\bigcup_{i=1}^n (P_i\times\Gamma_i^*\times 2^\Lang)}$ as usual.

\medskip
Next, we consider a more general problem, single-indexed LTL model-checking problem for L-DPNs with regular valuations,
where the validity of each atomic proposition depends on the control locations, the set of locks and the stack content.
Since a local configuration consists of a control location, the stack content and a set of held locks, it is natural
to consider this model-checking problem. Formally, a regular valuation is a function $\lambda:  AP\longrightarrow 2^{\bigcup_{i=1}^n (P_i\times\Gamma_i^*\times 2^\Lang)}$ such that for every $ap\in AP$, the set of local configurations that satisfies $ap$ is a regular set of local configurations, i.e., there is a L-MA $M$ such that $(p\omega,L)\in \lambda(ap)$ iff $(p\omega,L,\emptyset)\in L(M_{ap})$.

\subsection{A Motivating Example}

\begin{figure}[t]
\begin{lstlisting}[mathescape,xleftmargin=0.02\textwidth,frame=single, language==pascal, basicstyle=\small, multicols=2,columns=flexible, numbers=left, numberstyle=\tiny,numbersep=5pt, stepnumber=1]
public class Main{
  public static void main(){
    Socket sSocket= new Socket(2013);
    Socket cSocket=null;
    while(cSocket=sSocket.accept()){
      Worker work = new Worker(cSocket);
	  work.run();
} } }

public class Worker extends Thread{}{
  Socket s=null;
  public Worker(Socket c)
  { s=c; }
  public void run(){
  String str=null;
  Resource res;
    while(str=s.readLine()){
      if(str==``q") break;
      else res=getResource(str);
ap$_1$:   synchronized(res){
ap$_2$:   //do some critical stuff on
        //res with s;
     }
ap$_3$://leave critical section
    }
    s.close();
} }
\end{lstlisting}
\caption{A Java-like concurrent server program. $a$, $b$ and $c$ atomic propositions associated with the corresponding control points.}
\label{motivated-example}
\end{figure}
Figure \ref{motivated-example} shows a simplified Java-like concurrent server program.
The main process creates a socket object \emph{sSocket} to listen on the port \emph{2013}. When a request (connection) arrives from a client, the main process creates a new process \emph{work}, passes the connection \emph{cSocket} to \emph{work}, executes the new process \emph{work} by invoking \emph{run()} and continues listening on the port \emph{2013}. The process \emph{work} accepts a request from
the main process and processes it. If it receives ``q" from the client, it closes the connection by invoking \emph{close()} and this process terminates. Otherwise, it obtains the resource \emph{res} according to the request \emph{str} and checks for availability of the lock $l$ implicitly associated with \emph{res} by \emph{synchronized(res)}. If $l$ is free, \emph{work} successfully acquires $l$, performs the critical operations, after that $l$ is released. Otherwise, if $l$ has already been acquired by another process \emph{work'}, then
\emph{work} becomes blocked until $l$ is free.  This program has an unbound number of processes of \emph{Worker} which communicate
with other processes via nested locks.

L-DPNs are well-suitable to model such kind of concurrent programs. Several interesting properties of this program can be expressed in single-indexed LTL formulas. Suppose the scheduler is fair. $ap_1$, $ap_2$ and $ap_3$ atomic propositions associated with the corresponding control points.  The \emph{starvation} property can be expressed as $\F ap_1\wedge \G \neg ap_2$ checking whether there is a local run of \emph{Worker} running in parallel in the network such that at some point, $ap_1$ holds but $ap_2$ will never hold in the future, i.e.,
the local run will never enter the critical section. The \emph{deadlock} property can be expressed as $\F \G ap_1$ checking
whether or not the lock $l$ can be held forever by some process. The property that each process always leave the synchronized block
once it enters can be expressed as $\F ap_2\wedge \G\neg ap_3$.
The \emph{mutual exclusion} over the shared resource
can also be checked by our techniques.
We can create two DPDSs $\PP$ and $\PP'$ for \emph{Worker} in which the atomic proposition $ap_2$ is named by $ap_2'$ in $\PP'$. The main process alternatively creates new processes using
$\PP$ and $\PP'$ when a request arriving. The mutual exclusion property is expressed as $\F ap_2\wedge \F ap_2'$.

\subsection{Single-indexed LTL Model-Checking for Dynamic Pushdown Networks}
\label{LTL-DPN}
Single-index LTL model-checking for DPNs with simple and regular valuations was studied  in \cite{ST13dpn}.
We recall these results in this section, since we will
reduce single-indexed LTL model-checking for L-DPNs to these problems.
Let $\pi(p)$ denote the index $i$ such that $p\in P_i$.

\medskip
\begin{theorem}(Thm. 3 of \cite{ST13dpn})
\label{ch-dpn-DCLIC-fixpoint}We can compute a set $\CC_{fp}$ of DCLICs in time ${\bf O}(\sum_{p\omega\in\CC_I}(|\omega|\cdot |\delta_{\pi(p)}|\cdot |Q_{\pi(p)}|)\cdot 2^{|\CC_I|}+|\CC_I|^2\cdot 2^{|\CC_I|})$ s.t. for every $c\in\CC_I$, $c$ satisfies the single-indexed LTL formula $f$ iff $c\in\CC_{fp}$.
\end{theorem}

\begin{theorem} (Thm. 4 of \cite{ST13dpn})
\label{ch-dpn-single-indexed-LTL} Given a DPN $\MM=\{\PP_1,...,\PP_n\}$, a single-indexed LTL
formula $f=\bigwedge_{i=1}^n f_i$ and a simple function $\lambda$,
we can compute MAs  $\AAA_1,...,\AAA_n$ in time $\textit{O}(\sum_{i=1}^n(|\Delta_i|\cdot2^{|f_i|}\cdot|\Gamma_i|\cdot |P_i|^3)\cdot 2^{|\CC_I|})$ s.t. for every global configuration $p\omega$, $p\omega$
satisfies $f$ iff there exists $D\subseteq \CC_{fp}$  s.t. $(p\omega,D)\in L(\AAA_{\pi(p)})$.
\end{theorem}

\begin{theorem}
\label{single-indexed-LTL-regular-valuations}(Thm. 5 of \cite{ST13dpn})
Given a DPN $\MM=\{\PP_1,...,\PP_n\}$, a single-indexed LTL
formula $f=\bigwedge_{i=1}^n f_i$ and a regular valuation $\lambda$,
we can compute MAs $\AAA_1,...,\AAA_n$ in time ${\bf O}(\sum_{i=1}^n(|\Delta_i|\cdot2^{|f_i|}\cdot|\Gamma_i|\cdot|States_i|\cdot |P_i|^3)\cdot 2^{|\CC_I|})$
s.t. for every global configuration $p\omega$, $p\omega$
satisfies $f$ iff there exists $D\subseteq \CC_{fp}$  s.t. $(p\omega,D)\in L(\AAA_{\pi(p)})$, where $|States_i|$ denotes the number of states of the automata corresponding to the regular valuation $\lambda$.
\end{theorem}

\section{Single-indexed LTL Model-Checking for L-DPNs}
\label{lock-elm}
To check whether a L-DPN $\MM$ satisfies $f$ is non-trivial, we cannot directly apply the approach of \cite{ST13dpn} to check whether the L-DPN $\MM$ satisfies $f$ or not, as we have to
ensure that the access of locks in each instance is correctly coordinated with other instances.
To solve this problem, we follow the work of \cite{LMW09} which reduces reachability checking of L-DPNs to checking reachability of DPNS,
We will reduce single-indexed LTL model-checking for L-DPNs to single-indexed LTL model-checking for DPNs.
For this, for every $i,1\leq i\leq n$, we compute a new DPDS
$\PP'_i$, which is a kind of ``product" of the DPDS $\PP_i$ with acquisition structures, where an
acquisition structure contains lock usage information such as the set of held locks, the order dependence of acquisition and release of
locks. We will associate each control location of $\PP_i'$ with an acquisition structure. The acquisition structure stored in a local configuration ``guesses" the acquisition and release histories of locks in the global run starting from this local configuration.
From the  acquisition structures, we can infer the lock usages of a global run.
Then, we can obtain a DPN $\MM'=(\PP_1',...,\PP_n')$ such that the global runs of $\MM'$ mimic the global runs of $\MM$. Since we can check lock usages from the acquisition structures, we disallow all the acquisition structures that violate the nested lock style in the global run of $\MM'$. Thus, the global runs of $\MM'$ exactly correspond to the global runs of $\MM$ that use locks in a nested style.
We can get that $\MM$ satisfies $f$ iff $\MM'$ satisfies $f$.
The later problem can be solved by Theorems \ref{ch-dpn-DCLIC-fixpoint}, \ref{ch-dpn-single-indexed-LTL} and \ref{single-indexed-LTL-regular-valuations}.

\subsection{Acquisition Structures}
\label{acq-str}
Along a global run $T$ of $\MM$,  a release of a lock $l$ without a corresponding acquisition of $l$
in the same local run is called \emph{initial release}. An acquisition of a lock $l$ without a corresponding release of $l$
in the same local run is called \emph{final acquisition}.
An \emph{acquisition structure}  $as$ of a global run $T$ is a tuple $(R, RH, U,AH,A,X)$, where $R$ (resp. $A$) is the set of initial release (resp. final acquisition) locks of $T$;
$U$ is a set of \emph{usages}, i.e., acquisition and release locks that are not final acquisition or initial release locks are called;  $RH\subseteq \Lang\times\Lang$ is a \emph{release graph} such that
$(l,l')\in RH$ iff $T$ has an initial release of $l'$ and the usages of $l$
occurs before the initial release of $l'$;
$AH\subseteq \Lang\times\Lang$ is an \emph{acquisition graph} such that $(l,l')\in AH$ iff $T$ has a final acquisition of $l$
and the usages of $l'$ occurs after the final acquisition of $l$; $X$ is a set of locks that
are initially-held at the root.

An acquisition structure $as=(R, RH, U,AH,A,X)$ is \emph{consistent} iff both $RH$ and $AH$ are acyclic, and
$(X\setminus R)\cap (U\cup A)=\emptyset$. Intuitively, the set of locks $(X\setminus R)$ denotes all the
initially-held locks that will not be released during the run.
Thus, these locks cannot be used (i.e., $(X\setminus R)\cap U=\emptyset$) or finally acquired (i.e., $(X\setminus R)\cap A=\emptyset$) anymore. The fact that the graphs $RH$ and $AH$ are acyclic ensures that the acquisition and release of locks do not have any cycle dependence. Intuitively, if $RH$
has edges $(l_1,l_2),...,(l_m,l_{m+1})$ for some $m>1$ such that $l_1=l_m{+1}$ (i.e., $RH$ has a cycle), then for every $i:1< i\leq m+1$,
the lock $l_i$ has an initial release in the global run $T$ and should be performed after a usage of
$l_{i-1}$ (according to the definition of release graphs). Since the initial release of $l_i$ releases the initially-held lock $l_i\in X$, then before releasing this initially-held lock $l_i$, $l_i$ cannot be acquired anymore, i.e., the usage of $l_i$ occurs after the initial release of $l_i$.  Thus, the initial release of $l_{i+1}$ should be done after the initial release of $l_{i}$.
Since $l_1=l_{m+1}$, then, the initial release of $l_1$ should be performed after the initial release of
$l_1$. We deduce a deadlock. Thus, $RH$ should be acyclic. $AH$ is similar.
Let $AS$ be the set of all the consistent acquisition structures.
Given an acquisition structure $as=(R, RH, U,AH,A,X)$, we will write $as_R=R$, $as_{RH}=RH$,
$as_U=U$, $as_{AH}=AH$, $as_A=A$ and $as_{X}=X$.

\begin{example}
Let us consider the global run $T$ given in Figure \ref{nested-locks}(II).
During the local run $n_1(n_2n_6n_7n_8)^*$, it will always create a new instance whose local run is $n_3n_4n_5$.
Let $(R, RH, U,AH,A,X)$ be the acquisition structure of the node $n_1$ in $T$.
Suppose the set of initially-held locks in $T$ is $\emptyset$, then $X=\emptyset$.
Since all the releases in $T$ has an corresponding acquisition, we can know that
$R=\emptyset$ which implies that $RH=\emptyset$. While the acquisition of the locks except $acq(l_1)$
does not have any corresponding release, thus $A=\{l_1\}$. The acquisition of the locks $l_2$ and $l_3$ (i.e.
$acq(l_2), acq(l_3)$) have corresponding releases (i.e., $rel(l_2), rel(l_3))$, we deduce that
$U=\{b,c\}$. The order of the final acquisition of $l_1$ and the usages of $l_2$ and $l_3$ gives us that
$AH=\{(l_1,l_2),(l_1,l_3)\}$. Similarly,
the acquisition structure of $T^{n_6}$ (i.e., the subtree rooted by the node $n_6$) is
$(R', RH', U',AH',A',X')$, where $R'=\{l_2\}$ (since the initial release $rel(l_2)$ does not have any corresponding acquisition of $l_2$), $X'=\{l_1,l_2\}$ (the locks $l_1$ and $l_2$ acquired when moving from $n_1$ to $n_6$ are not yet released), $RH'=\{(l_3,l_2)\}$, $A'=AH'=\emptyset$, $U'=\{l_2,l_3\}$.
\end{example}

Given two consistent acquisition structures $as=(R_1, RH_1, U_1,AH_1,A_1,X_1)$ and
$as'=(R_2, RH_2, U_2,AH_2,A_2,X_2)$, $as$ and $as'$ are \emph{compatible}, denoted by $Compatible(as,as')$, iff the following conditions hold:

\begin{center}
\begin{tabular}{lcl}
1. $X_1\cap X_2=\emptyset$; & & 2. $(A_1\cup (X_1\setminus R_1))\cap (A_2\cup (X_2\setminus R_2))=\emptyset$;\\
3. $RH_1\cup RH_2$ is acyclic; & & 4. $AH_1\cup AH_2$ is acyclic;\\
5. $(A_1\cup U_1)\cap (X_2\setminus R_2)=\emptyset$  && 6. $(A_2\cup U_2)\cap (X_1\setminus R_1)=\emptyset$.
  \end{tabular}
\end{center}

Roughly specking, the compatible condition is used to check whether two global runs
could be two subtrees of a node in a global run.
Imagine there is a global run such that $(p\omega,L)$ is a leaf and $(p_1\omega_1,L_1)$
(resp. $(p_2\omega_2,L_2)$) could be the right (resp. left) child of the leaf $(p\omega,L)$ using a transition rule with $\tau$ action.
Suppose $T_1$ and $T_2$ be two global runs that are rooted by $(p_1\omega_1,L_1)$ and $(p_2\omega_2,L_2)$, respectively. Let $as^1$ and $as^2$ be the acquisition structures of $T_1$ and $T_2$, respectively. The $compatible(as^1,as^2)$ checks whether
the two tree $T_1$ and $T_2$ could be two subtrees of the leaf $(p\omega,L)$.
Each condition verifies whether the nested lock access is violated or not. Items 1 and 2 verify
that the initially-held locks and finally-held hocks are disjoint. If there is a lock $l\in X_1\cap X_2$, then
the lock $l$ will be held by the two local configuration $(p_1\omega,L_1)$ and  $(p_2\omega_2,L_2)$.
This implies that $T_1$ and $T_2$ cannot be the subtrees of the node $(p\omega,L)$ in the global run of $\MM$ due to a deadlock. On the other hand, if  there is a lock $l\in (A_1\cup (X_1\setminus R_1))\cap (A_2\cup (X_2\setminus R_2))$, then, $T_1$ (resp. $T_2$) has a final acquisition of the lock $l$ or
$l$ is held at the initial global configuration of $T_1$ (resp. $T_2$) that will not be released. This means that
both $T_1$ and $T_2$ will finally hold the lock $l$ which is a deadlock.
Items 3 and 4 verify that the acquisition and release graphs are acyclic. Since if there is a cycle in $RH_1\cup RH_2$ or $AH_1\cup AH_2$,
as discussed previously, it will have a deadlock.
Items 5 and 6 verify that the held throughout locks (the set of initially-held locks that will not be
released) are not acquired or released.  If there is a lock $l\in (A_1\cup U_1)\cap (X_2\setminus R_2)$, then,
the lock $l$ will always be held in $T_2$. This implies that $T_1$ should not acquire $l_1$ (i.e, $l_1\not\in A_1\cup U_1$).

\subsection{From L-DPN to DPN}
In this section, we show how to compute a DPN  $\MM'=(\PP_1',...,\PP_n')$, a kind of ``product" of the L-DPN $\MM$ with the
acquisition structures $AS$ such that  a global run $T'$ of $\MM'$ mimics a corresponding global run $T$ of $\MM$, i.e.,
for every node $(p\omega,L)$ in $T$, there is a corresponding node $(p,as)\omega$ in $T'$, where
$as$ is the acquisition structure of the subtree rooted by $(p\omega,L)$ in $T$.
We update the acquisition structures embedded in the control locations of $\PP_i'$ during the global run of $\MM'$ and checks
whether the acquisition structure is consistent or not.
If an inconsistent acquisition structure occurs in a global run of $\MM'$, then the corresponding global run of $\MM$
violates the lock usages. Thus, we disallow all the global runs of $\MM'$ in which an inconsistent acquisition structure occurs.
Then, $\MM$ has a global run $T$ that starts from a global configuration $(p\omega,L)$, uses locks in a nested style and satisfies $f$ iff $\MM'$ has a corresponding global run $T'$ starting from a global configuration $(p,as)\omega$ for some $as\in AS$ such that $T'$ satisfies $f$.

To compute $\MM'$, for every $i: 1\leq i\leq n$, let $\PP_i'=(P_i',\Gamma_i,\Delta_i')$,
where $P_i'=P_i\times AS$ and $\Delta_i'$ is computed as follows: for every $as, as',as''\in AS$,

\begin{enumerate}
  \item $(p,as)\gamma\stackrel{\tau}{\hookrightarrow}_i (p',as')\omega\in\Delta_i'$  iff $p\gamma\stackrel{a}{\hookrightarrow}_i p'\omega\in\Delta_i$, and one of the following conditions holds:
        \medskip
      \begin{enumerate}
        \item[1.1:] $as=as'$,  if $a=\tau$;  or
        \medskip
        \item[1.2:] $as=\Big(as_R'\cup\{l\}, as_{RH}', as_U', as_{AH}',as_A',as_{X}'\cup \{l\}\Big)$, if $a=rel(l)$ and $l\not\in as_X'\cup as_R'$; or
                \medskip
        \item[1.3:] $as=\left\{\begin{array}{l}
                           \Big(as_R'\setminus\{l\}, (as_{RH}'\setminus(\Lang\times\{l\})\cup (\{l\}\times as_R'\setminus\{l\})), as_U'\cup\{l\}, as_{AH}',as_A',\\ as_{X}'\setminus\{l\}\Big),
                            \hfill{\mbox{if } a=acq(l)\mbox{ and }  l\in as_R'\cap as_X';}  \\
                              \\
                           \Big(as_R', as_{RH}', as_{U}', as_{AH}'\cup\{\{l\}\times as_U'\},as_A'\cup\{l\},as_{X}'\setminus\{l\}\Big),\\
                           \hfill{\mbox{else if } a=acq(l)\mbox{ and }  l\not\in as_A'\mbox{ and } l\in as_X';}
                         \end{array} \right.$
      \end{enumerate}
 \medskip
   \item $(p,as)\gamma\stackrel{\tau}{\hookrightarrow}_i (p',as')\omega\rhd (p_1,as'')\omega_1\in\Delta_i'$  iff $p\gamma\stackrel{a}{\hookrightarrow}_i p'\omega\rhd p_1\omega_1\in\Delta_i$,  $Compatible(as',as'')$, $as''_R=as''_X=\emptyset$,
       and  one of the following conditions holds:
         \medskip
      \begin{enumerate}
        \item[2.1:] $as=(as'_R\cup as_R'', as'_{RH}\cup as_{RH}'', as_{U}'\cup as_{U}'',as_{AH}'\cup as_{AH}'',as_{A}'\cup as_{A}'',as_{X}'\cup as_{X}'')$,  if $a=\tau$; or
        \medskip
        \item[2.2:] $as= \Big(as'_R\cup as''_R\cup\{l\}, as'_{RH}\cup as''_{RH}, as'_U\cup as''_U, as'_{AH}\cup as''_{AH},as'_A\cup as''_A, as'_X\cup as''_{X}\cup \{l\}\Big)$,  if  $a=rel(l)$ and $l\not\in as'_X\cup as''_X\cup as'_R\cup as''_R$;          or

            \medskip
        \item[2.3:] $as=\left\{\begin{array}{l}
                           \Big((as'_R\cup as_R'')\setminus\{l\}, ((as'_{RH}\cup as_{RH}'')\setminus(\Lang\times\{l\})\cup (\{l\}\times (as'_R\cup as_R'')\setminus\{l\})), \\
                           as'_U\cup as_U''\cup\{l\}, as'_{AH}\cup as_{AH}'',as'_A\cup as_A'',(as'_X\cup as_X'')\setminus\{l\}\Big),\\
                            \hfill{\mbox{if }  a=acq(l)\mbox{ and } l\in (as'_R\cup as_R'')\cap (as'_X\cup as_X'');}  \\
                             \\
                           \Big(as'_R\cup as_R'', as'_{RH}\cup as_{RH}'', as'_U\cup as_U'', \{\{l\}\times (as'_U\cup as_U'')\} \cup as'_{AH}\cup as_{AH}'',\\
                           as'_A\cup as_A''\cup\{l\},(as'_X\cup as_X'')\setminus\{l\}\Big), \\
                           \hfill{\mbox{else if }  a=acq(l)\mbox{ and }  l\not\in as'_A\cup as_A'' \mbox{ and } l\in as'_X\cup as_X''.}
                         \end{array} \right.$
      \end{enumerate}
 \end{enumerate}

Intuitively, the DPN $\MM'=(\PP_1',...,\PP_n'$ is a kind of ``product" of $\MM$ with acquisition structures $AS$
such that $\MM$ has a global run $T$ starting from a global configuration $(p\omega,L)$ such that $T$ uses locks in a nested style and satisfies $f$ iff
$\MM'$ has a corresponding global run $T'$ starting from a global configuration $(p,as)\omega$ for some $as\in AS$ such that $T'$ satisfies $f$. Intuitively, if we  reserve  only the set of held locks $X$ of each acquisition structure in the global runs of $\MM'$,
then the global runs of $\MM'$ are exactly the global runs of $\MM$.
The acquisition structure updated during the global runs of $\MM'$ ensures that
the consistent acquisition structure $as$ of the root $(p,as)\omega$ is the
acquisition structure of the tree $T$, i.e., $T$ uses locks in a nested style.
Let us explain the intuition behind each items by structural induction on the global run of $\MM$ ($\MM'$).
Suppose the global run $T$ of $\MM$ has a node $(p\omega,L)\in P_i\times\Gamma_i^*\times 2^\Lang$ s.t.
$\omega=\gamma u$ for some $i: 1\leq i\leq n$, $\gamma\in\Gamma_i$ and
the subtree rooted by $(p\omega,L)$ is $T_1$.

If the node $(p\omega,L)$ has the right and left
children $(p'v u,L)$ and $(p_2\omega_2,\emptyset)$ due to the transition rule $p\gamma\stackrel{\tau}{\hookrightarrow}_i p'v\rhd p_2\omega_2\in\Delta_i$, then we suppose that the subtree rooted by $(p'v u,L)$ (resp. $(p_2\omega_2,\emptyset)$) in $T$ be $T_2$
(resp. $T_3$) and the acquisition structure of $T_2$ (resp. $T_3$) is $as'$ (resp. $as''$).
According to the definition of the acquisition structures, $as''_X=\emptyset$. This implies that
$as''_R=\emptyset$, since there does not any lock need to release.
If $Compatible(as',as'')$ (as explained in Section \ref{acq-str}) does not hold or $as'$ (resp. $as''$) is inconsistent, then the global run $T$ violates the lock usages. This implies
that all the global runs of $\MM'$ should not contain a node  $(p,as')'v u$ (resp. $(p_2,as'')\omega_2)$). Thus, we do not add
a transition rule $(p,as)\gamma\stackrel{\tau}{\hookrightarrow}_i (p',as')\omega\rhd (p_1,as'')\omega_1$ into $\Delta_i'$
for any $as\in AS$ if $Compatible(as',as'')$ does not hold, or $as'$ (resp. $as''$) is inconsistent.
Otherwise, according to the definition of acquisition structures,
the acquisition structure $as$ of $T_1$ must be $(as'_R\cup as_R'', as'_{RH}\cup as_{RH}'', as_{U}'\cup as_{U}'',as_{AH}'\cup as_{AH}'',as_{A}'\cup as_{A}'',as_{X}'\cup as_{X}'')$.
This implies that if $(p,as)\gamma v$ is a node of a global run of $\MM'$, then $(p',as')u v$ and $(p_1,as'')\omega_1$
can be the right and left children of $(p,as)\gamma v$. For this, we add the transition rule $(p,as)\gamma\stackrel{\tau}{\hookrightarrow}_i (p',as')\omega\rhd (p_1,as'')\omega_1$ into $\Delta_i'$.

If the node $(p\omega,L)$ has the right and left
children $(p'v u,L\setminus\{l\})$ and $(p_2\omega_2,\emptyset)$ due to the transition rule $p\gamma\stackrel{rel(l)}{\hookrightarrow}_i p'v\rhd p_2\omega_2\in\Delta_i$, then $l\in L$ (see Item $\alpha_6$).
We suppose that the subtree rooted by $(p'v u,L)$ (resp. $(p_2\omega_2,\emptyset)$) in $T$ be $T_2$
(resp. $T_3$) and the acquisition structure of $T_2$ (resp. $T_3$) is $as'$ (resp. $as''$), then
$l\not\in as'_X$, $as''_X=\emptyset$, $as''_X=\emptyset$ and
$as''_R=\emptyset$ according to the definition
of the acquisition structures. This implies that $T_2$ and $T_3$ should not have any initial release of $l$, (i.e., $l\not\in as'_R\cup as''_R$). Thus, if $l\in as'_X\cup as''_X\cup as'_R\cup as''_R$, we do not add
a transition rule $(p,as)\gamma\stackrel{\tau}{\hookrightarrow}_i (p',as')\omega\rhd (p_1,as'')\omega_1$ into $\Delta_i'$
for any $as\in AS$.
Otherwise, if $Compatible(as',as'')$ does not hold or $as'$ (resp. $as''$) is inconsistent, then  we neither do not add
a transition rule $(p,as)\gamma\stackrel{\tau}{\hookrightarrow}_i (p',as')\omega\rhd (p_1,as'')\omega_1$ into $\Delta_i'$
for any $as\in AS$ as explained above.  If $l\not\in as'_X\cup as''_X\cup as'_R\cup as''_R$
and $Compatible(as',as'')$ holds, according to the definition of acquisition structures,
the acquisition structure $as$ of $T_1$ must be $\Big(as'_R\cup as''_R\cup\{l\}, as'_{RH}\cup as''_{RH}, as'_U\cup as''_U, as'_{AH}\cup as''_{AH},as'_A\cup as''_A, as'_X\cup as''_{X}\cup \{l\}\Big)$ (note that $rel(l)$
is an initial release for $T_1$). This implies that if $(p,as)\gamma v$ is a node of a global run of $\MM'$, then $(p',as')u v$ and $(p_1,as'')\omega_1$ can be the right and left children of $(p,as)\gamma v$. For this, we add the transition rule $(p,as)\gamma\stackrel{\tau}{\hookrightarrow}_i (p',as')\omega\rhd (p_1,as'')\omega_1$ into $\Delta_i'$.

If the node $(p\omega,L)$ has the right and left
children $(p'v u,L\cup\{l\})$ and $(p_2\omega_2,\emptyset)$ due to the transition rule $p\gamma\stackrel{acq(l)}{\hookrightarrow}_i p'v\rhd p_2\omega_2\in\Delta_i$, then $l\not\in L$ (see Item $\alpha_4$).
We suppose that the subtree rooted by $(p'v u,L)$ (resp. $(p_2\omega_2,\emptyset)$) in $T$ be $T_2$
(resp. $T_3$) and the acquisition structure of $T_2$ (resp. $T_3$) is $as'$ (resp. $as''$), then
$l\in as'_X$, $as''_X=\emptyset$ and $as''_R=\emptyset$ according to the definition
of the acquisition structures. Thus, if $l\not\in as'_X\cup as_X''$, $Compatible(as',as'')$ does not hold,
$as''_X\neq\emptyset$, $as''_R\not\emptyset$ or $as'$ (resp. $as''$) is inconsistent, then, the global run $T$
violates the lock usages. On the other hand, if $l\not\in as'_R$ and $l\in as_A'\cup as_A''$,
then the lock $l$ acquired by applying $p\gamma\stackrel{acq(l)}{\hookrightarrow}_i p'v\rhd p_2\omega_2$
will finally acquired again before the release of $l$. This arises a deadlock.
Thus, we do not add $(p,as)\gamma\stackrel{\tau}{\hookrightarrow}_i (p',as')\omega\rhd (p_1,as'')\omega_1$ into $\Delta_i'$
for any $as\in AS$ if ($l\not\in as'_R$ and $l\in as_A'\cup as_A''$), or $l\not\in as'_X\cup as_X''$, or $Compatible(as',as'')$ does not hold,
or $as''_X\neq\emptyset$, or $as''_R\not\emptyset$ or $as'$ (resp. $as''$) is inconsistent.

Otherwise, if $l\in as'_R$, $Compatible(as',as'')$, $as'$ and $as''$ are consistent,
according to the definition
of the acquisition structures, the acquisition structure $as$
of $T_1$ must be $\Big((as'_R\cup as_R'')\setminus\{l\}, ((as'_{RH}\cup as_{RH}'')\setminus(\Lang\times\{l\})\cup (\{l\}\times (as'_R\cup as_R'')\setminus\{l\})), as'_U\cup as_U''\cup\{l\}, as'_{AH}\cup as_{AH}'',as'_A\cup as_A'',(as'_X\cup as_X'')\setminus\{l\}\Big)$.
Indeed, the initial release of $l$ together with $acq(l)$ will be a usage of $l$ for $T_1$ instead of the initial release of $l$ which implies
that $as_{AH}$ should not contain any edge $\Lang\times\{l\}$. This usage of $l$ occurs after the other initial releases $as_R'\setminus\{l\}$. Thus, if $(p,as)\gamma v$ is a node of a global run of $\MM'$, then $(p',as')u v$ and $(p_1,as'')\omega_1$ can be the right and left children of $(p,as)\gamma v$. For this, we add the transition rule $(p,as)\gamma\stackrel{\tau}{\hookrightarrow}_i (p',as')\omega\rhd (p_1,as'')\omega_1$ into $\Delta_i'$.

Otherwise, if $l\not\in as'_R$, $l\in as_A'\cup as_A''$, $Compatible(as',as'')$, $as'$ and $as''$ are consistent,
then, according to the definition
of the acquisition structures, the acquisition structure $as$
of $T_1$ must be $\Big(as'_R\cup as_R'', as'_{RH}\cup as_{RH}'', as'_U\cup as_U'', \{\{l\}\times (as'_U\cup as_U'')\} \cup as'_{AH}\cup as_{AH}'', as'_A\cup as_A''\cup\{l\},(as'_X\cup as_X'')\setminus\{l\}\Big)$.
Indeed, $acq(l)$ is a final acquisition of $l$ due to $l\not\in as'_R$ and the usages $as'_U\cup as_U''$
occur after this final acquisition of $l$. Thus, if $(p,as)\gamma v$ is a node of a global run of $\MM'$, then $(p',as')u v$ and $(p_1,as'')\omega_1$ can be the right and left children of $(p,as)\gamma v$. For this, we add the transition rule $(p,as)\gamma\stackrel{\tau}{\hookrightarrow}_i (p',as')\omega\rhd (p_1,as'')\omega_1$ into $\Delta_i'$.

 \medskip
Thus, we can get that  $\MM$ has a global run $T$ starting from a global configuration $(p\omega,L)$ using locks in a nested style
iff $\MM'$ has a corresponding global run $T'$ starting from a global configuration $(p,as)\omega$ for some $as\in AS$.
By extending the valuation $\lambda:AP\longrightarrow 2^{\bigcup_{i=1}^n (P_i\times\Gamma_i^*\times 2^\Lang)}$
to $\lambda':AP\longrightarrow 2^{\bigcup_{i=1}^n (P_i\times AS\times\Gamma_i^*)}$,
we can get the following theorem. Indeed, for every $ap\in AP$,  $\lambda'(ap)=\{ (p,as)\omega \in \bigcup_{i=1}^n (P_i\times AS\times\Gamma_i^*) \mid (p\omega,as_X)\in \lambda(ap)\}$.

\begin{theorem}
\label{decomthm-2}
Given a L-DPN $\MM=(Act,\Lang,\PP_1,...,\PP_n$ s.t. for every $i$,
$1\leq i\leq n$, $\PP_i=(P_i,\Gamma_i,\Delta_i)$, a LTL formula $f=\bigwedge_{i=1}^n f_i$ and a valuation $\lambda$,
let $\MM'=(\PP_1',...,\PP_n')$ be the DPN
such that for every $i:1\leq i\leq n$, the DPDS $\PP_i'$ is computed as above. For every global configuration
$(p_0\omega_0,\emptyset)$, $\MM$ has a global run $T$ starting from
$(p_0\omega_0,\emptyset)$ that uses locks in a nested style and satisfies $f$ iff $\MM'$ has a global run $T'$ starting from $(p_0,as)\omega_0$ for some $as\in AS$ such that $T'$ satisfies $f$. Moreover,
for every $i:1\leq i\leq n$, $|P_i'|={\bf O}(|P_i|\cdot 2^{{\bf O}(|\Lang|^2)})$, $|\Delta_i'|={\bf O}(|\Delta_i|\cdot 2^{{\bf O}(|\Lang|^2)})$, and $\MM'$ can be computed in time ${\bf O}(\sum_{i=1}^n|\Delta_i|\cdot 2^{{\bf O}(|\Lang|^2)})$.
\end{theorem}

The complexity follows from the fact that the number of acquisition graphs $AH$ and release graphs $RH$
is at most $2^{{\bf O}(|\Lang|^2)}$ which implies that the number acquisition structures is at most  $2^{{\bf O}(|\Lang|^2)}$.

\medskip
By Theorem \ref{decomthm-2}, Theorem \ref{ch-dpn-single-indexed-LTL} and Theorem \ref{single-indexed-LTL-regular-valuations}, we can get
the following two theorems. Note that the number of DCLICs in $\MM'$ is at most $|\CC_I|\cdot 2^{\textit{O}(|\Lang|^2)}$.

\begin{theorem}
\label{L-DPN-LTL}Given a L-DPN $\MM=(Act,\Lang,\PP_1,...,\PP_n$, a single-indexed LTL
formula $f=\bigwedge_{i=1}^n f_i$ and a simple function $\lambda$,
we can compute MAs  $\AAA_1,...,\AAA_n$ in time $\textit{O}(\sum_{i=1}^n(|\Delta_i|\cdot2^{|f_i|}\cdot|\Gamma_i|\cdot |P_i|^3)\cdot 2^{\textit{O}(|\Lang|^2)} \cdot 2^{|\CC_I|}\cdot 2^{\textit{O}(|\Lang|^2)})$ s.t. for every $i: 1\leq i\leq n$ and every $(p\omega,L)\in P_i\times\Gamma_i^*\times 2^\Lang$, $(p\omega,L)$ satisfies $f$ iff there exist $D\subseteq \CC_{fp}$ and $as\in AS$ s.t. $((p,as)\omega,D)\in L(\AAA_{\pi(p)})$.
\end{theorem}

\begin{theorem}
\label{L-DPN-LTL-regular-valuations}
Given a L-DPN $\MM=(Act,\Lang,\PP_1,...,\PP_n$, a single-indexed LTL
formula $f=\bigwedge_{i=1}^n f_i$ and a regular valuation $\lambda$,
we can compute MAs $\AAA_1,...,\AAA_n$ in time ${\bf O}(\sum_{i=1}^n(|\Delta_i|\cdot2^{|f_i|}\cdot|\Gamma_i|\cdot|States_i|\cdot |P_i|^3)\cdot 2^{\textit{O}(|\Lang|^2)} \cdot 2^{|\CC_I|\cdot 2^{\textit{O}(|\Lang|^2)} })$
s.t. for every $i: 1\leq i\leq n$ and every $(p\omega,L)\in P_i\times\Gamma_i^*\times 2^\Lang$, $(p\omega,L)$ satisfies $f$ iff there exist $D\subseteq \CC_{fp}$ and $as\in AS$ s.t. $((p,as)\omega,D)\in L(\AAA_{\pi(p)})$ and $as_X=L$, where $|States_i|$ denotes the number of states of the automata corresponding to the regular valuation $\lambda$.
\end{theorem}

\section{Related work}
\label{rel}
{\bf DPNs and L-DPNs:} The DPN model was introduced in \cite{BOT05}.
Several other works use DPN and its extensions to model multi-threaded programs \cite{BOT05,GLOSW11,LOW09,Lug11,Wen10}. All these works only consider reachability issues.
Ground Tree Rewrite Systems \cite{GL11} and process
rewrite systems \cite{BKRS09,Mayr00} are two models of multi-threaded programs
with procedure calls and threads creation. However, \cite{Mayr00} only considers reachability problem and \cite{GL11,BKRS09}
only consider subclasses of LTL.  We consider single-indexed LTL model checking problems.

Pushdown networks with communication between processes are studied in \cite{BET03,CCKRT06,ABT08,TA10,CMV12}.
These works consider systems with a fixed number of threads. \cite{LO07,LO08} use parallel flow graphs to model multi-threaded programs.
However, all these works only consider reachability. \cite{Yah01} considers safety properties of multi-threaded programs.

\medskip
\noindent
{\bf Lock usages and acquisition structures:} \cite{KIG05} first introduces (forward) acquisition histories that contains only the set of held locks and the acquisition graphs
to check pairwise reachability properties of two threads (i.e. pushdown systems) communicating via well-nested locks.
\cite{KG06,KG07} extended the results of \cite{KIG05} with backwards acquisition histories that contains only the set of held locks and the release graphs to check fragments of LTL and CTL properties for two threads communicating via well-nested locks.
\cite{LO08} extended the acquisition histories of \cite{KIG05} to check pairwise reachability
properties of programs with reentrant monitors (a restricted form of well-nested locks) and dynamic thread creation.
\cite{KLTR09} uses tuples-of-lock histories to check
pushdown networks without threads creation for properties
represented by a kind of finite automata.  Using tuples-of-lock histories allows
the decision procedure to use only one reachability query of each pushdown system. While, in the worst case, \cite{KIG05} has to
perform an exponential number of individual reachability queries of each pushdown system to handle an temporal operator.
In order to compute predecessor sets of regular sets of configurations of L-DPNs,
\cite{LMW09} introduces acquisition structures for L-DPNs. Their acquisition structures are similar to tuples-of-lock histories of
\cite{KLTR09} which are defined for pushdown networks without threads creation.
\cite{LMW09} reduces the predecessor sets computation of L-DPNs to compute the predecessor sets of of DPNs \cite{BOT05}.

Following \cite{LMW09}, in this work, we reduce single-indexed LTL model-checking for L-DPNs to
single-indexed LTL model-checking for DPNs. This latter problem can be solved by our previous work \cite{ST13dpn}.
In this work, the acquisition structures we used are similar to the acquisition structures of \cite{LMW09}.
However, in \cite{LMW09}, the acquisition structures do not contain the set of held locks (i.e., $as_X$).
and \cite{LMW09} first encodes the acquisition structures into a hedge automaton, a kind of finite automaton, and then
computes the product DPN of the L-DPN with the hedge automaton, where the set of held locks are stored
into the control locations of the DPN. Our work do not perform this intermediate procedure, i.e.,
encoding the acquisition structures into a hedge automaton. Instead, we directly compute the product DPN from a
L-DPN and the acquisition structures. Our approach can remove inconsistent  acquisition structures during this product.
While, \cite{LMW09} removes them when computing the product DPN of the L-DPN and the acquisition structures.
Moreover, during encoding of the acquisition structures into a hedge automaton,
\cite{LMW09} checks only whether the sets of finally acquired locks of two acquisition structures are disjoint.
when composing two threads. This may introduces some inconsistent acquisition structures (e.g., having cycles in a release graph) into the hedge automaton. While, in this work, we check the compatibility of two acquisition structures
which can remove these inconsistent acquisition structures. This makes the DPN more small.

Recently, \cite{Kah09} introduces bounded lock chains, a generalization of well-nested locks and shows that pairwise reachability is decidable for pushdown networks (without threads creation) with bounded lock chains.
\cite{Kah11} extends the results of \cite{Kah09} to show the decidability of the fragment LTL for
pushdown networks without threads creation.
\cite{CMV12} introduces contextual locking, another extension of well-nested locks, and shows that pairwise reachability for pushdown networks (without threads creation) with contextual locking. These works do not consider threads creation.
It is unknown whether our approach checks single-indexed LTL properties for DPNs with lock chains or contextual locking.
We leave them as future work.

\bibliographystyle{abbrv}
\bibliography{bib}
\end{document}